\documentclass[12pt,preprint]{aastex}
\begin{document}

\title{The Non-Damped Nature of Twelve
 Low-Redshift Damped Ly$\alpha$ Candidate 
Systems\altaffilmark{1}}

\author{David A. Turnshek\altaffilmark{2}
and Sandhya M. Rao\altaffilmark{2}}

\affil{Department of Physics \& Astronomy, University of Pittsburgh,
Pittsburgh, PA 15260, USA}

\affil{Accepted for Publication in the Astrophysical Journal Letters}

\altaffiltext{1}{Based on data obtained with the NASA/ESA 
{\it Hubble Space Telescope} at the Space Telescope Science Institute, 
which is operated by AURA, Inc., under NASA contract NAS5-26555.}

\altaffiltext{2}{email: turnshek@quasar.phyast.pitt.edu,
rao@everest.phyast.pitt.edu}

\begin{abstract}
Hubble Space Telescope ($HST$)
UV spectroscopy of 12 candidate low-redshift damped Ly$\alpha$ (DLA)
systems in 11 QSOs ($z=0.103$ in Q0054+144, $z=0.969$ and $z=0.987$ in Q0302$-$223,
$z=0.478$ in Q0454$-$220, $z=1.476$ in Q1047+550, $z=1.070$ in Q1206+459,
$z=1.228$ in Q1247+267, $z=0.399$ in Q1318+290B, $z=0.519$ in Q1329+412,
$z=0.276$ in Q1451$-$375, $z=0.204$ in Q2112+059, $z=0.263$ in Q2251+113)
are presented; the observations demonstrate that they are not DLAs with
$N_{HI}\ge2\times10^{20}$ atoms cm$^{-2}$. In all cases except two the
systems either do not exist or are well below the DLA threshold column
density; the exceptions are a $z=0.474$ system in Q0454$-$220 which has
$N_{HI}=3\times10^{19}$ atoms cm$^{-2}$ and a $z=1.223$ system in
Q1247+267 which has $N_{HI}=8\times10^{19}$ atoms cm$^{-2}$.  
Despite the availability
of data in the $HST$ archives demonstrating that these are not suitable
targets, many have unfortunately been approved for observation with $Chandra$, 
$Gemini$, and/or $HST$ with the intent of doing followup work on low-redshift 
DLAs. Furthermore, these results indicate that the low-redshift DLA statistics 
derived from {\it IUE} spectra and presented by Lanzetta, Wolfe, \& Turnshek (1995)
and Wolfe et al. (1995) are invalid.
\end{abstract}


\

\

\section{INTRODUCTION}

Damped Ly$\alpha$ (DLA) QSO absorption line systems trace the neutral
gas in the non-local Universe. Consequently, studies of DLAs are
used to investigate many fundamental problems in galaxy formation
research, from measurements of the evolution of the cosmological
mass density and extent of the neutral gas component, to studies
of the types of galactic reservoirs that harbor the neutral gas, to
measurements of the neutral-gas-phase cosmic metallicity. However,
despite recent successes in identifying DLAs at low redshifts (Rao \&
Turnshek 2000, hereafter RT2000), they are still relatively rare,
with fewer than 30 now known at redshifts $z<1.65$ corresponding to
the wavelength where the Ly$\alpha$ line falls in the UV shortward
of the earth's atmospheric transmission cutoff.

Unfortunately, part of the extragalactic astronomy community has recently
been confused about the reality of some low-redshift DLA systems.
Initial confusion probably stemmed from a lack of distinction between
candidate DLA systems, identified by various means, and ones that have
been confirmed through UV spectroscopy with adequate resolution and
signal-to-noise ratio. In particular, the International Ultraviolet
Explorer ($IUE$) study of Lanzetta, Wolfe, \& Turnshek (1995, hereafter
LWT95) presented a list of low-redshift {\it candidate} DLA systems, most
of which were shown not to be DLAs in followup $HST$ UV spectroscopy.
Evidently the status of these systems needs to be clarified because
investigators are still being awarded time to make followup imaging and
metallicity studies of them on $Chandra$, $Gemini$, and $HST$. Moreover,
numerous researchers have cited and often used statistical results on
low-redshift DLAs that are based on these non-existent DLA systems. Thus,
our motivation for writing this short contribution is twofold. First, we
wish to inform observers and referees about which candidate low-redshift
DLA systems are not confirmed, so as to minimize the possibility that
valuable telescope time will be wasted in the future. Second, we wish
to correct the record on statistical results which make use of these
non-DLA systems.  Attempts have been made to clarify the situation in
at least two previous refereed papers (Januzzi et al. 1998; RT2000),
however those papers did not present the $HST$ spectra which prove that
the systems are not DLAs, although such evidence was easily accessible in
the $HST$ archives. In this paper we are more explicit.

The remainder of this paper is therefore broken into two parts.
In the first part (\S2) we present results on LWT95 candidate DLA
systems which have been shown not to be damped based on followup $HST$
UV spectroscopy, and we cite some examples of wasted telescope time
allocated to observe them. In the second part (\S3) we discuss statistical
results on low-redshift DLA systems which have been derived using these
non-systems, and we identify some prominent results in the literature
which make use of or quote these erroneous results.

\section{The Non-Damped Low-Redshift DLA Candidate Systems}

For technical reasons having to do with the ability to detect
DLA systems in the initial survey for them, the study of Wolfe et
al. (1986) defined them as those systems with neutral hydrogen column
densities $N_{HI}\ge2\times10^{20}$ atoms cm$^{-2}$; this threshold
corresponds to a Ly$\alpha$ absorption line on the ``damping'' part of
the curve-of-growth with a rest equivalent width (REW) of 10 \AA.  At the
time this definition was motivated by instrumental limitations, since
systems with lower $N_{HI}$ could not be reliably detected.  Thereafter,
all surveys for large column density neutral hydrogen systems, and the
``DLA'' jargon describing them, have adopted this same threshold. In
part this is undoubtedly because it was realized that surveys at this
threshold capture the bulk of the neutral gas in the Universe out to at
least redshift $z<3.5$. Since the incidence of Ly$\alpha$ absorption
systems rises with decreasing $N_{HI}$ for at least $N_{HI}>10^{13}$
atoms cm$^{-2}$, it is important to take this threshold into account in
any statistical survey.  At the same time, the relative rarity of the
highest column density systems introduces a statistical uncertainty that
is not easy to assess, and there is some worry that the highest $N_{HI}$
systems which are metal-rich and have high dust content could be missed in
magnitude-limited surveys due to dimming of the background QSO.  But the
highest $N_{HI}$ systems are the ones that dominate certain observational
determinations of cosmological interest for constraining the galaxy
formation process, for example, the mean cosmological mass density of
the neutral gas component or the cosmic neutral-gas-phase metallicity,
which is a column-density-weighted measurement. Therefore, for example,
in these cosmological determinations of the properties of the neutral
gas, the discovery and measurement of a high column density system with
$N_{HI}=2\times10^{21}$ atoms cm$^{-2}$ will have as much importance or
weight as ten DLA systems at the threshold column density value.

The $IUE$ study of LWT95 was the first to present an analysis of a large
number of UV spectra with the aim of identifying candidate DLA systems
at low redshifts ($z<1.65$). Approximately 260 individual $IUE$ spectra
were considered in the study.  In the final analysis 16 candidate DLA
systems were identified in the spectra of 14 QSOs. To be conservative
LWT95 included suspected Ly$\alpha$ absorption lines with REWs as low
as 5 \AA\ in the candidate DLA list, allowing for the possibility that
better data may reveal some of the weaker ones to be true DLA systems with
REW$\ge10$ \AA. Of the 16 candidate systems, four could immediately be
ruled out as DLA absorbers (LWT95), leaving the nature of the remaining
12 to be determined using future $HST$ UV spectroscopy. Four of these 12
had REW$\ge10$ \AA, and eight had $10>$REW$\ge5$ \AA.

$HST$ UV followup spectroscopy is available in the $HST$ archives for
three of the four LWT95 candidate DLA systems with REW$\ge10$ \AA, and
one has been confirmed as a DLA system ($z_{abs}=1.372$ in Q0935+417,
Januzzi et al. 1998). The one with REW$\ge10$ \AA\ that has not been
observed with $HST$ (the $z_{abs}=0.484$ candidate DLA in Q2223$-$052,
also called 3C446) was observed by Miller \& French (1978) and their
spectrum does not show the presence of a
\ion{Mg}{2}$\lambda\lambda$2796,2803 absorption doublet nor a
\ion{Fe}{2}$\lambda2600$ absorption line at the positions predicted by
the candidate system.  Based on this information and the RT2000
empirical MgII-FeII selection criterion for DLAs (i.e.,
DLAs have strong MgII-FeII absorption), it appears
highly unlikely that there is a DLA system at this redshift.
Chengalur \& Kanekar (2000) have also looked for 21 cm absorption at
the candidate redshift and did not find any.

$HST$ UV followup spectroscopy is available in the $HST$ archives for
seven of the eight weaker ($10>$REW$\ge5$ \AA) LWT95 candidate DLA
systems.  One has been confirmed as a DLA system ($z_{abs}=1.010$ in
Q0302$-$223, Pettini \& Bowen 1997, Boisse et al. 1998, RT2000). The
one candidate system which has not been observed in the UV with $HST$
($z_{abs}=1.342$ in Q0215+015) actually has a strong MgII-FeII system
at $z_{abs}=1.345$ with \ion{Mg}{2} $\lambda2796$ REW=1.88 \AA\ and
\ion{Fe}{2} $\lambda2600$ REW=1.45 \AA\ (Blades et
al. 1982). According to the RT2000 selection criterion for classical
DLAs, this system has an $\approx50$\% empirical probability of being
a DLA; however, this is the highest signal-to-noise ratio $IUE$
spectrum used in the LWT95 study and it suggests $N_{HI}=
8\times10^{19}$ atoms cm$^{-2}$, which is also consistent with the
findings of RT2000.

Thus, while two of the original 16 LWT95 candidate DLA systems do not
have $HST$ UV spectroscopy, the evidence suggest that it is unlikely
that they are DLAs.  The status of all of the LWT95 candidate
low-redshift DLA systems is shown in Table 1, along with comments
about the assumed reality of the systems at the time of the LWT95
publication. Figure 1 presents the available $HST$ UV spectroscopy for
the {\it non-damped} candidates; the wavelength locations of the {\it
non-damped} candidates are marked with arrows.

Despite these results, most of which have been available in the $HST$
archives as early as 1996, along with comments in several papers
generally pointing out the non-damped nature of many of the LWT95
candidate systems (e.g., Januzzi et al. 1998, RT2000, Fynbo, Moller,
\& Thomsen 2001, Bechtold et al. 2001), $Chandra$, $Gemini$, and $HST$
time have been awarded to followup several of them.  For example, a
$Chandra$ Guest Observer (GO) program  to study DLA metallicities
included the Q0054+144 candidate DLA in the target list, and $Chandra$
observed it.  A $Gemini$ program was allocated time to perform
adaptive-optics imaging of DLA galaxies, and several of the LWT95
non-damped systems were approved targets and were observed, including
the Q2251+113 field; the LWT95 paper itself pointed out that while
Q2251+113 formally had a candidate DLA system, an existing $HST$
spectrum showed that a DLA line was not present.  Most recently, $HST$
Cycle 11 time was awarded to a GO program  in Phase 1 to make
spectroscopic metallicity measurements of three of the LWT95 candidate
DLA systems, namely the ones reported in Q0054+144, Q1329+412, and
Q2112+059.  The Q2112+059 $HST$ observation was later dropped in Phase
2. However, as of March 2002, $HST$ STIS observations of the Q0054+144
and Q1329+412 DLA candidates are in the ``planning'' stages, even
though they do not exist.

\section{Implications for Low-Redshift DLA Statistics}

In addition to the issue of wasted telescope time, there is also the
issue of adoption of the LWT95 statistical result on DLAs at low
redshift. In the LWT95 study it was assumed that the four systems with
REW$\ge10$ \AA\ were true DLA systems with $N_{HI} \ge 2\times
10^{20}$ atoms cm$^{-2}$, while the eight systems with $10>$REW$\ge5$
\AA\ were not. These assumptions were used to plot the $z=0.8$ data
points in figures 3, 4, and 6 of LWT95.  However, as shown in \S2,
these were incorrect assumptions. The only system which LWT95 assumed
would be confirmed that actually was confirmed was a $z_{abs}=1.372$
DLA system in Q0935+417; a $z_{abs}=1.010$ DLA system in Q0302$-$223
was also confirmed, compensating for the others to some extent.
Nevertheless, these results invalidate the $z=0.8$ data point plotted
in figures 3, 4, and 6 of LWT95.  One might be tempted to try to infer
the real LWT95 low-redshift DLA statistics from these results, however
we believe this would be unwise since the redshift path reliably
surveyed by the $IUE$ spectra is clearly questionable.  We note that
shortly after the LWT95 study was complete, the LWT95 DLA low-redshift
statistics were adopted in the Wolfe et al. (1995) supplemental
high-redshift DLA survey.

Turnshek (1997), in a preview of the RT2000 results, first pointed
out that the often-cited low-redshift DLA results in LWT95 and Wolfe
et al. (1995) were invalid.  However, despite this and other concerns
which have appeared (Januzzi et al. 1998), many researchers continue
to use it, often to infer results on the evolution of DLA systems from
high to low redshift.  Two of the more relevant and prominent of the
investigations are Pei, Fall, \& Hauser (1999) and Storrie-Lombardi \&
Wolfe (2000), but for other investigations that may have used this result 
see citations to LWT95 and Wolfe et al. (1995) in the NASA Astrophysical
Data System Database. The $HST$ \ion{Mg}{2} survey results presented in
RT2000 are currently the only useful statistical results on
low-redshift DLAs that are available; unfortunately the error bars on
these results are currently quite large, but they are reliable.

\acknowledgements We thank the referee for his/her quick review,
suggestions, and encouragement. This work was supported in part by
STScI grant HST-GO-08569.01-A and NASA LTSA  grant NAG5-7930. This
research has made use of the NASA/IPAC Extragalactic  Database (NED)
which is operated by the Jet Propulsion Laboratory, California
Institute of Technology, under contract with the National Aeronautics
and Space Administration.

%\end{footnotesize}

\begin{deluxetable}{lccccccccr}
%\scriptsize
\rotate
\tablewidth{0pt}
\setlength{\tabcolsep}{1.5mm}
\tablenum{1}
\tablecaption{Status of Candidate DLA Systems from LWT95}
\tablehead{ 
\colhead{} & 
\colhead{LWT95} &
\colhead{LWT95} &
\colhead{LWT95} &
\colhead{LWT95} &
\colhead{Used in LWT95} &
\colhead{HST} &
\colhead{HST} &
\colhead{HST} &
\colhead{}\\[.2ex]
\colhead{QSO} & 
\colhead{$z_{abs}$} &
\colhead{Measured} &
\colhead{Candidate} &
\colhead{Inferred} &
\colhead{Low-z DLA} &
\colhead{Confirmed} &
\colhead{Confirmed} &
\colhead{Confirmed} &
\colhead{Notes}\\[.2ex]
\colhead{} &
\colhead{} &
\colhead{REW (\AA)} &
\colhead{DLA?} &
\colhead{$\log N_{HI}$} &
\colhead{Statistics?} &
\colhead{Status} &
\colhead{$z_{abs}$} &
\colhead{$\log N_{HI}$} &
\colhead{}
}

\startdata   
0054+144   & 0.1030 &  8.4 & Yes & 20.1    & No  & Not DLA & 0.1030 & 18.3 & 1 \\
0215+015   & 1.3420 &  6.4 & Yes & 19.9    & No  & See Note &\nodata &\nodata & 2 \\
0302$-$223 & 0.9690 &  5.7 & Yes & 20.0    & No  & Not DLA &\nodata &\nodata & 3 \\
\nodata    & 0.9874 &  5.6 & Yes & 20.0    & No  & Not DLA &\nodata &\nodata & 3 \\
\nodata    & 1.0140 &  6.7 & Yes & 20.0    & No  & DLA     & 1.0100 & 20.4 & 3\\
0454$-$220 & 0.4778 &  5.6 & No  & \nodata & No  & Not DLA & 0.4738 & 19.5 & 4 \\
0935+417   & 1.3691 & 10.3 & Yes & 20.3    & Yes & DLA     & 1.3720 & 20.5 & 5 \\
1047+550   & 1.4762 &  7.4 & Yes & 20.0    & No  & Not DLA &\nodata &\nodata &  6 \\
1206+459   & 1.0699 &  7.5 & No  & \nodata & No  & Not DLA &\nodata &\nodata & 7 \\
1247+267   & 1.2276 &  5.8 & Yes & 19.8    & No  & Not DLA & 1.2232 & 19.9 & 8 \\
1318+290B  & 0.3987 & 11.9 & No  & \nodata & No  & Not DLA &\nodata &\nodata & 9 \\
1329+412   & 0.5193 & 17.2 & Yes & 20.8    & Yes & Not DLA &\nodata &\nodata & 10 \\
1451$-$375 & 0.2761 &  7.9 & Yes & 20.1    & No  & Not DLA &\nodata &\nodata & 11 \\
2112+059   & 0.2039 & 11.5 & Yes & 20.4    & Yes & Not DLA &\nodata &\nodata & 12 \\
2223$-$052 & 0.4842 & 20.3 & Yes & 20.9    & Yes & See Note &\nodata &\nodata & 13 \\
2251+113   & 0.2633 &  9.0 & No  & \nodata & No  & Not DLA &\nodata &\nodata & 14 \\
\enddata
\vskip -0.3truein
\tablecomments{
(1) Bechtold et al. (2001); 
(2) No $HST$ spectrum available, Blades et al. (1982) find strong 
MgII-FeII absorption at $z=1.345$ but the high signal-to-noise ratio $IUE$ spectrum
gives $N_{HI}=8\times10^{19}$ atoms cm$^{-2}$;
(3) Pettini \& Bowen (1997), Boiss\'e et al. (1998), RT2000;
(4) $HST$ archive, PID 1026, M. Burbidge PI;
(5) Jannuzi et al. (1998);
(6) $HST$ archive, PID 5948, Lanzetta PI;
(7) Jannuzi et al. (1998);
(8) Pettini et al. (1999);
(9) Vanden Berk et al. (1999); 
(10) $HST$ archive, PID 5948, Lanzetta PI;
(11) $HST$ archive, PID 5948, Lanzetta PI;
(12) Fynbo et al. (2001);
(13) No $HST$ spectrum available, Miller \& French (1978) find no
MgII or FeII absorption in their optical spectrum,
non-detection at 21 cm (Chengalur \& Kanekar 2000);
(14) Bahcall et al. (1993).}

\end{deluxetable}

\clearpage

\begin{figure}
%\epsscale{0.2}
\plottwo{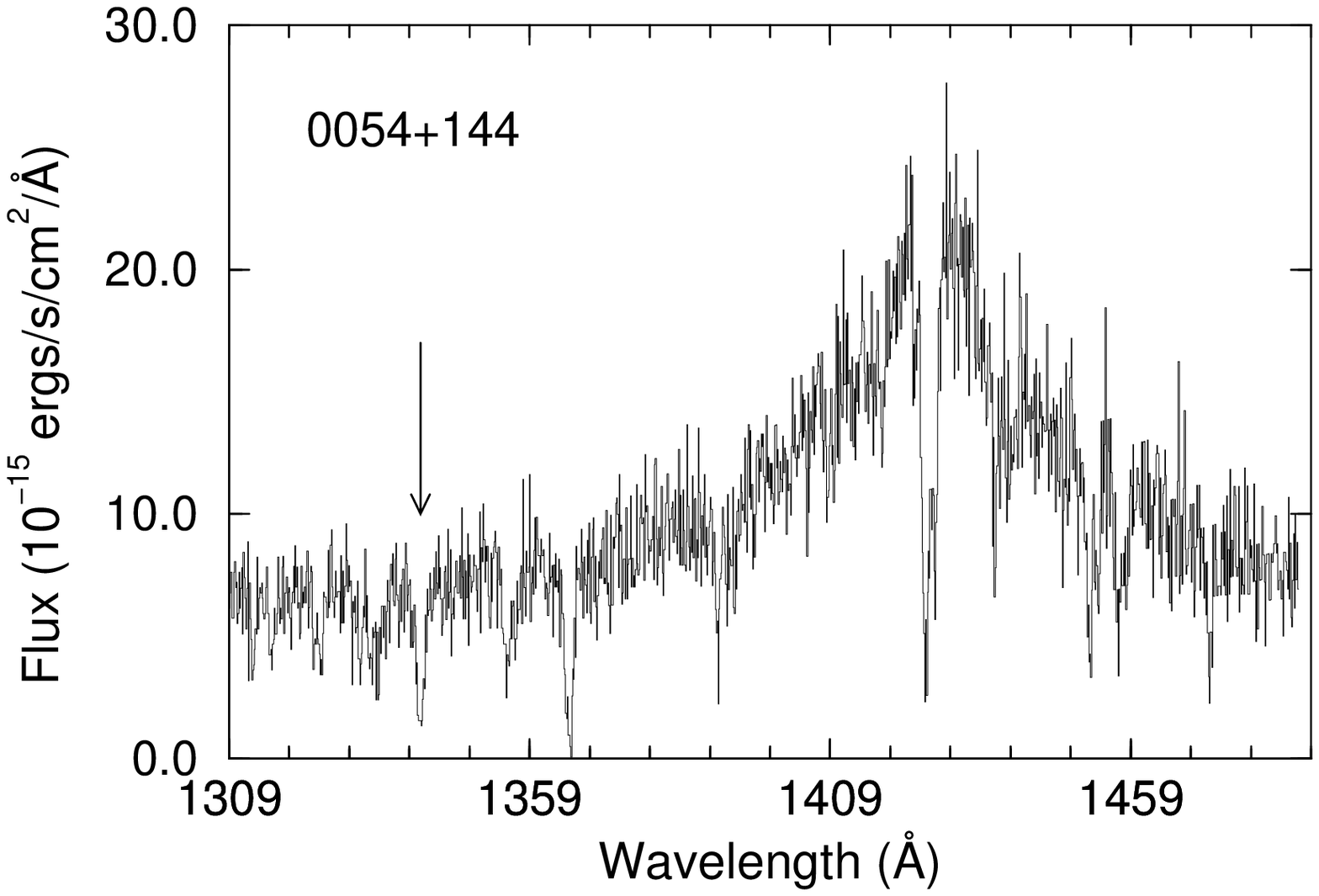}{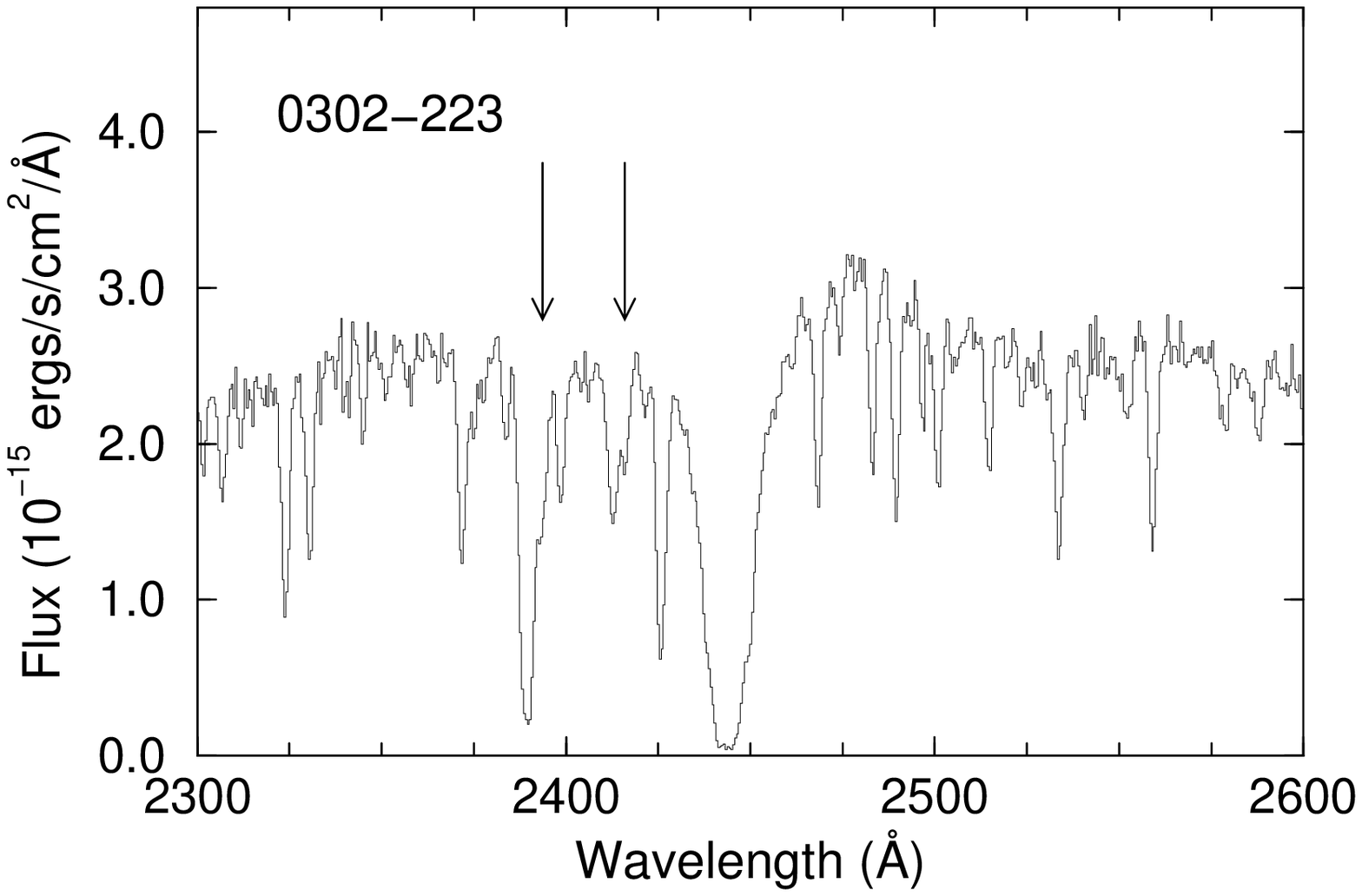}
\end{figure}
\begin{figure}
\plottwo{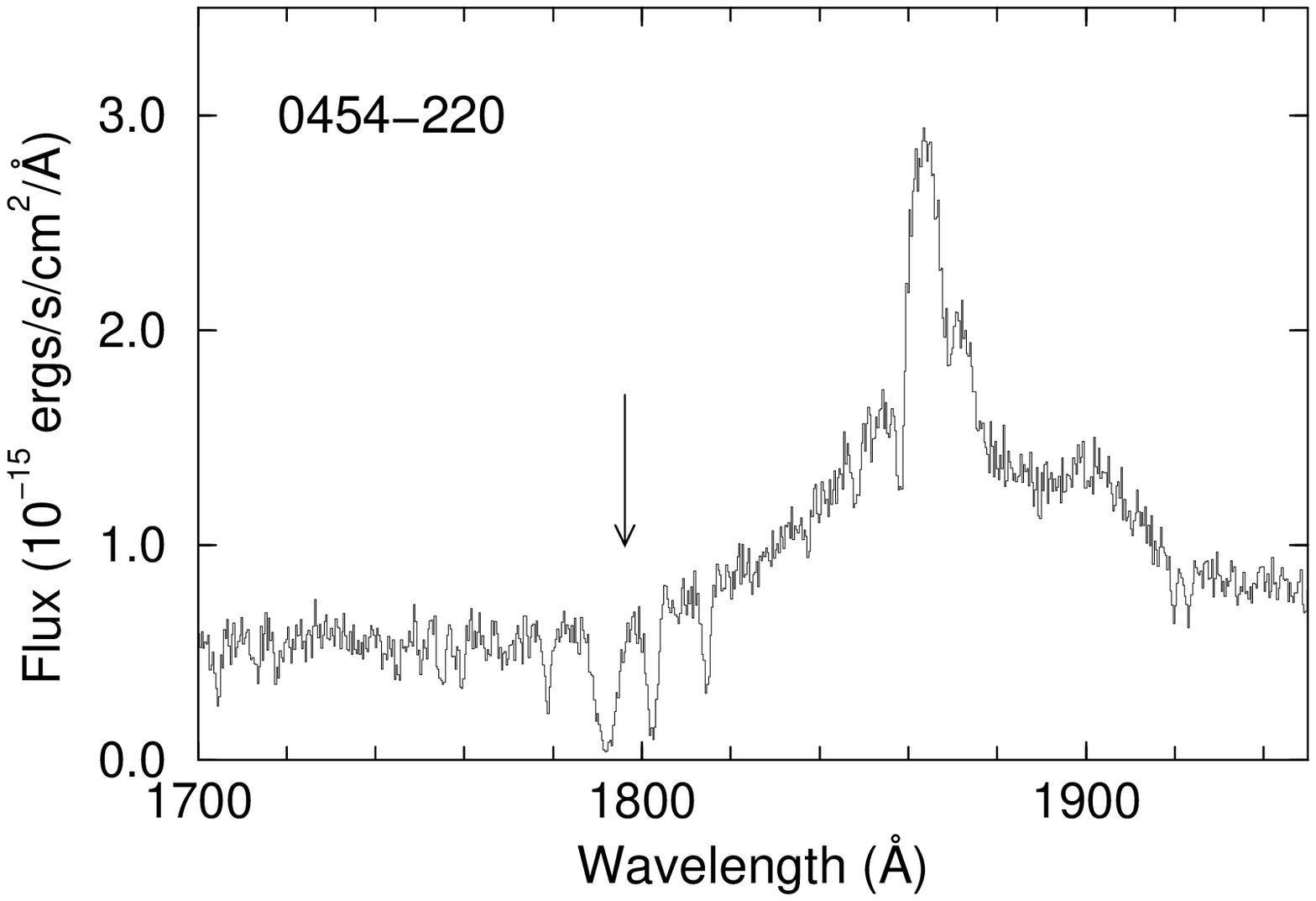}{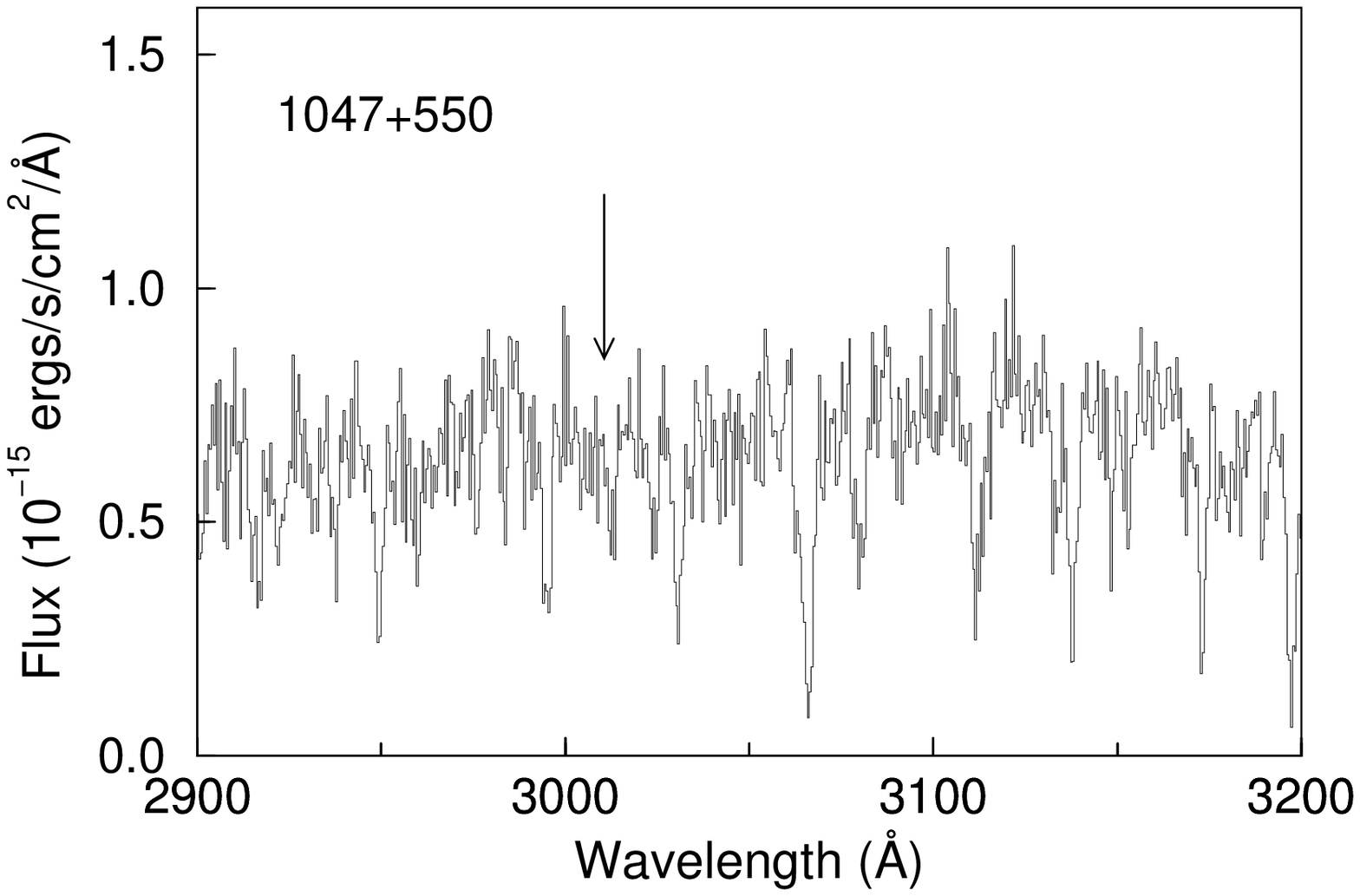}
\end{figure}
\begin{figure}
\figurenum{1}
\plottwo{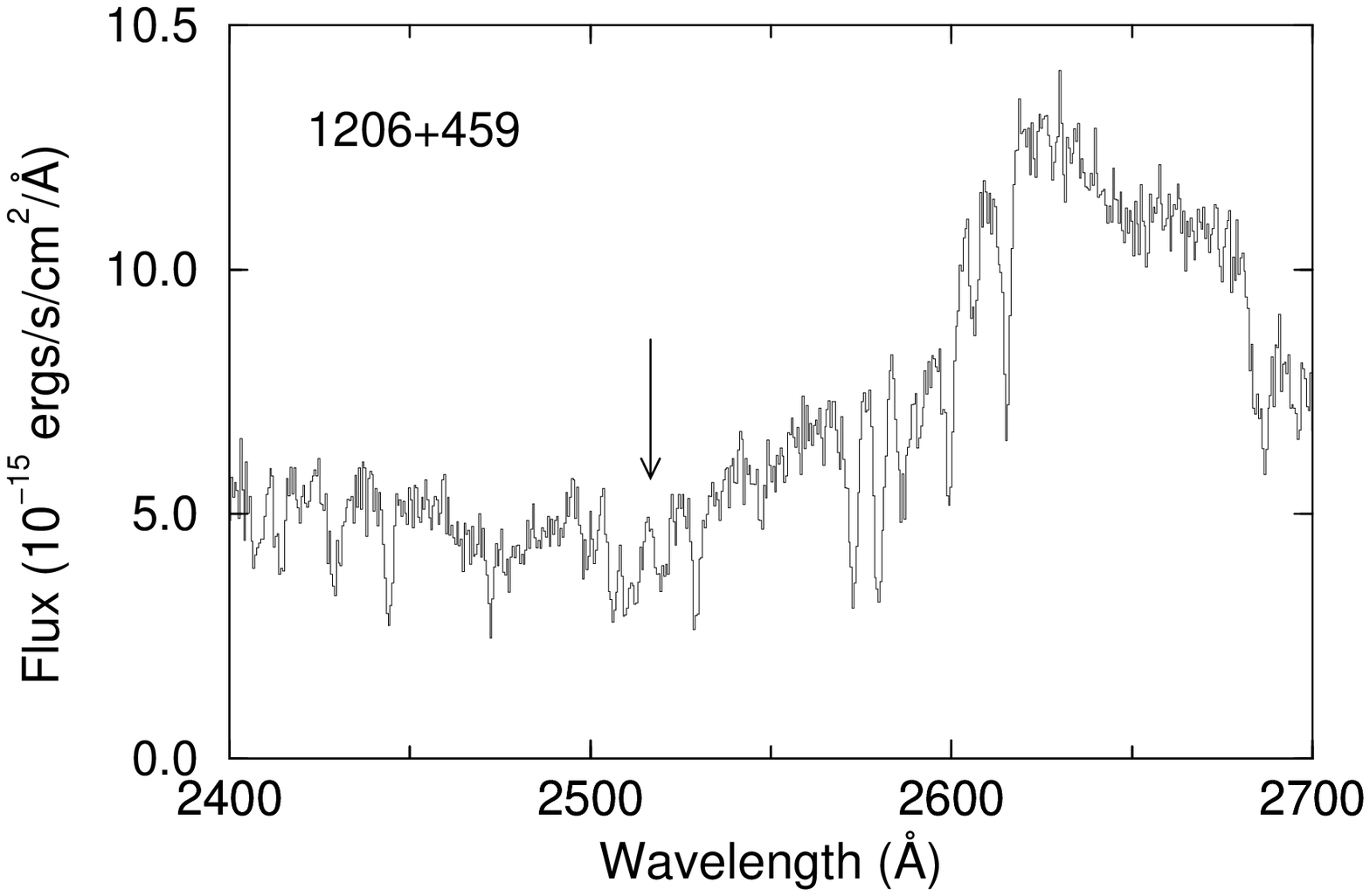}{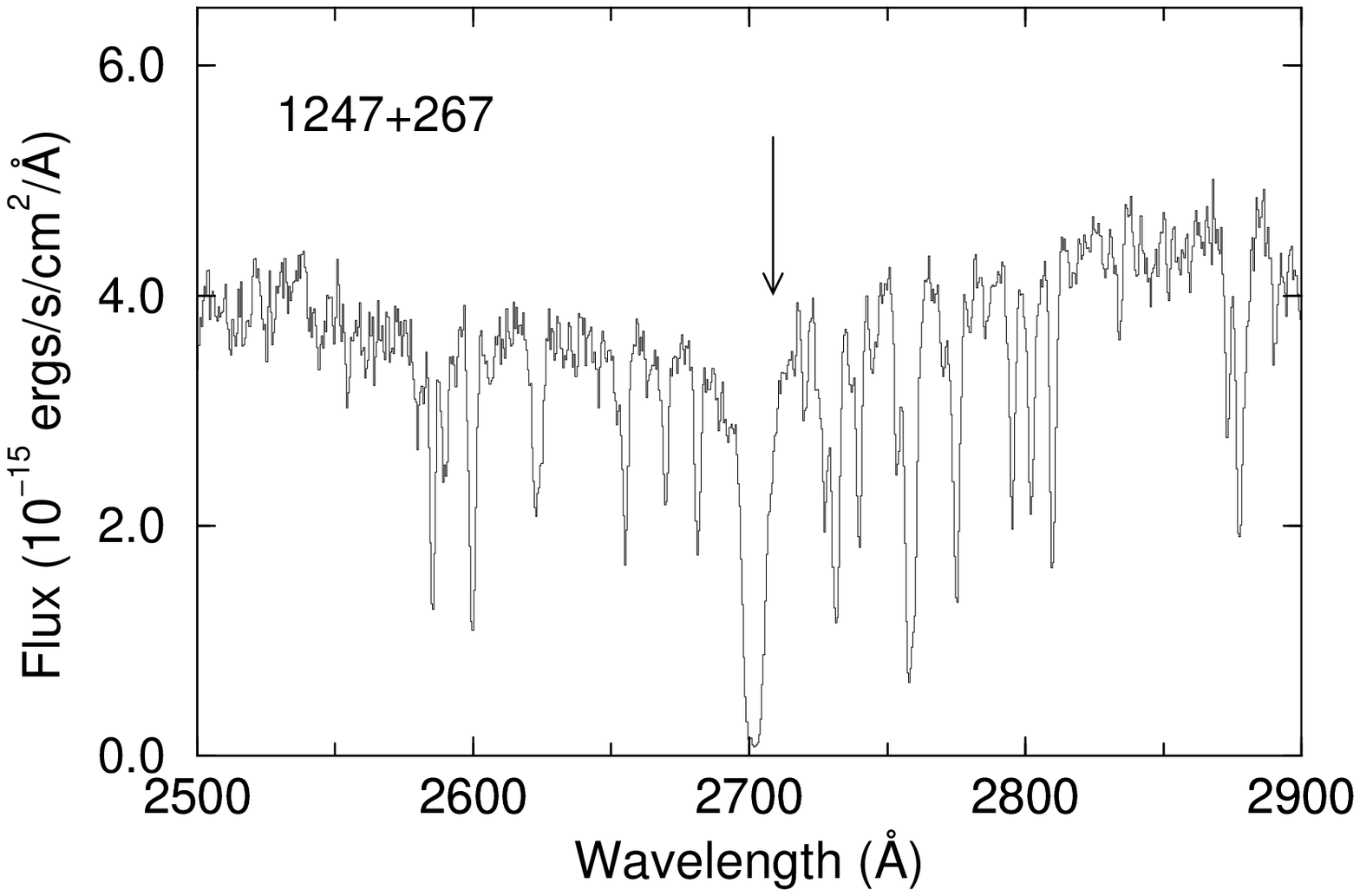}
\caption{HST UV spectra showing the absence of DLA lines
at the positions (marked by arrows) predicted by the 
LWT95 candidate DLA redshifts. The one actual DLA line
among these spectra is at 2444 \AA\ in Q0302$-$223 at $z=1.0100$.
Originally, this QSO had 3 DLA candidates, and the two 
candidates marked on the Q0302$-$223 spectrum are the ones 
ruled out.}
\end{figure}
\begin{figure}
\plottwo{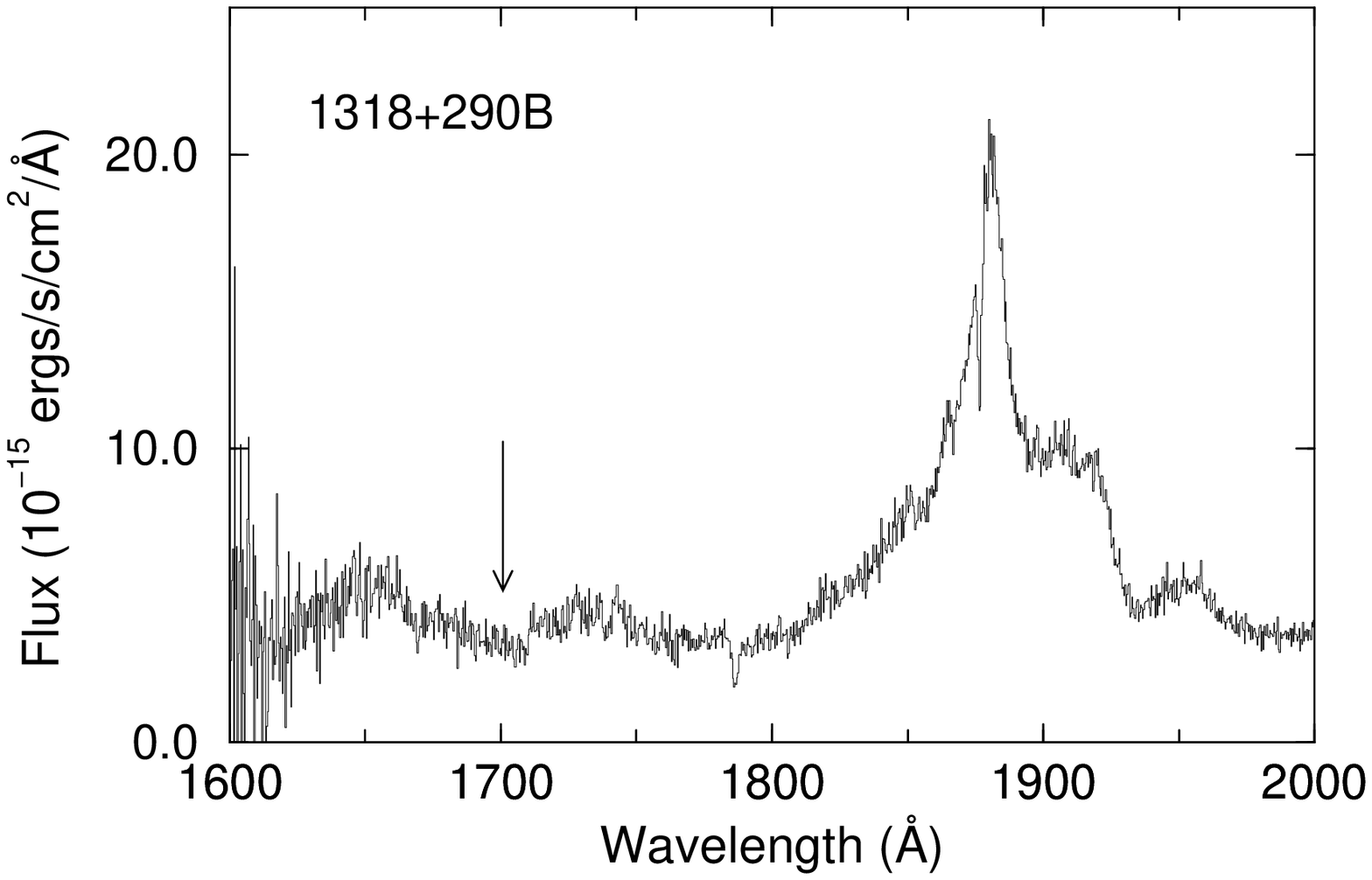}{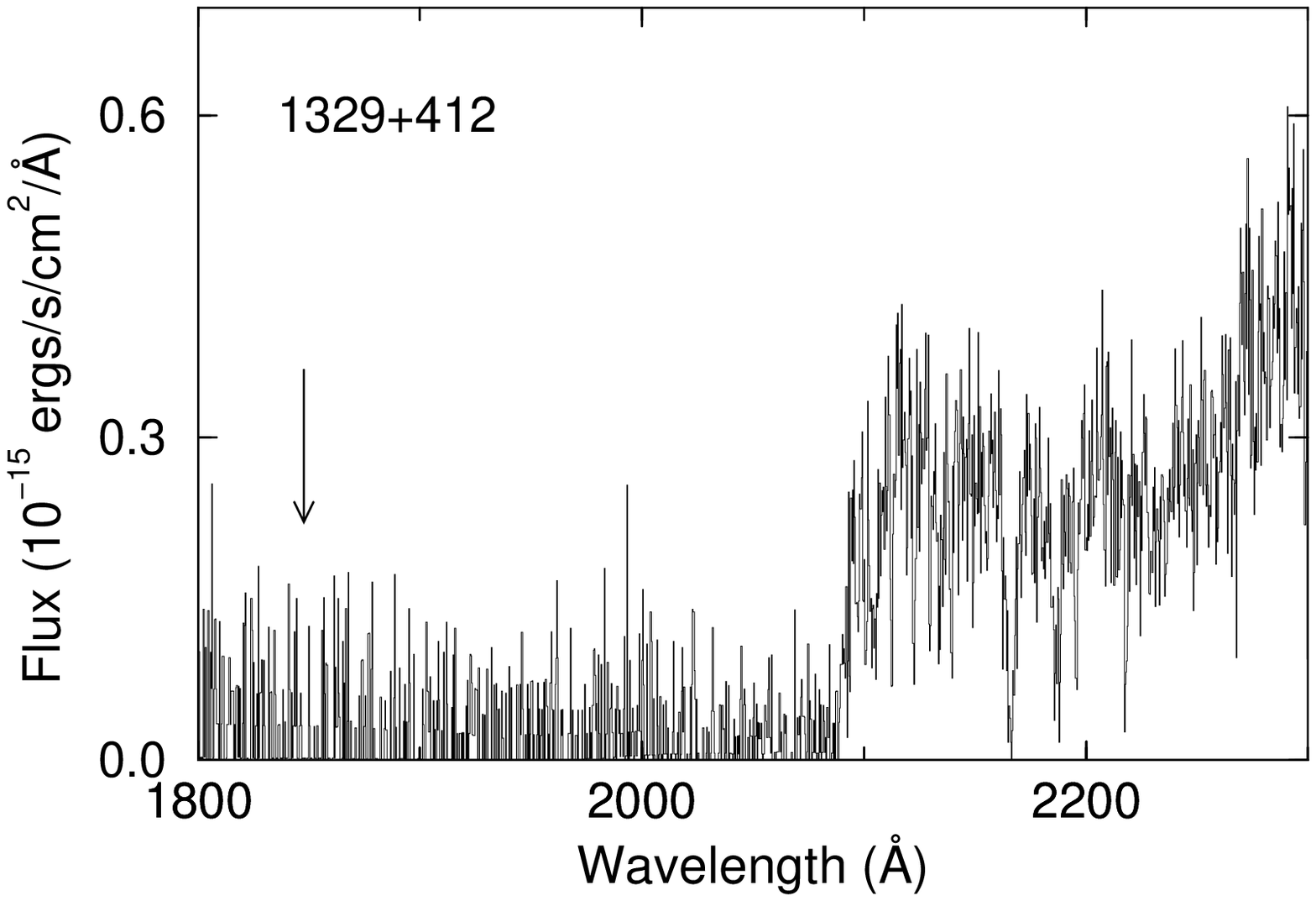}
\end{figure}
\begin{figure}
\plottwo{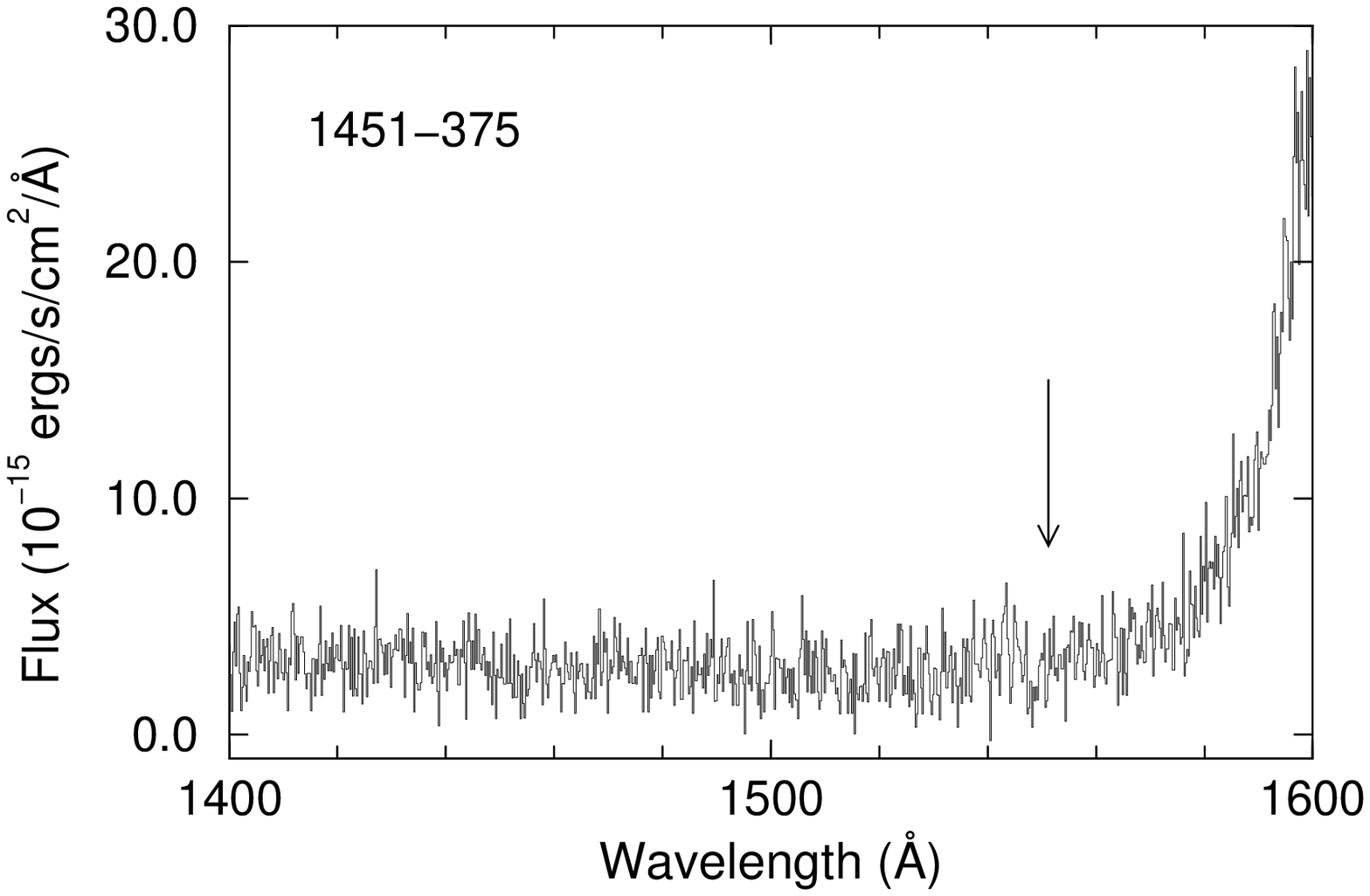}{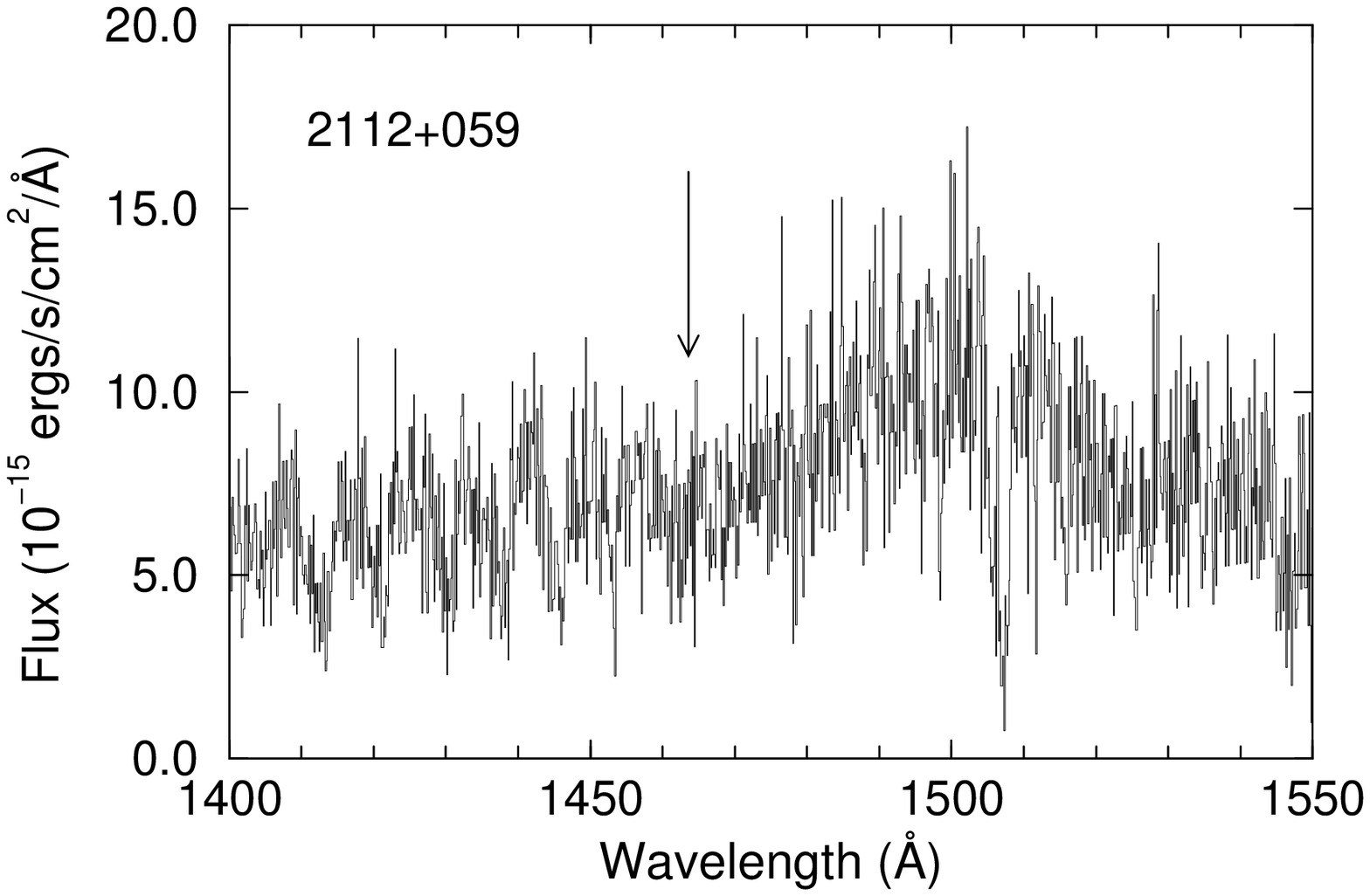}
\end{figure}
\begin{figure}
\figurenum{1}
\plottwo{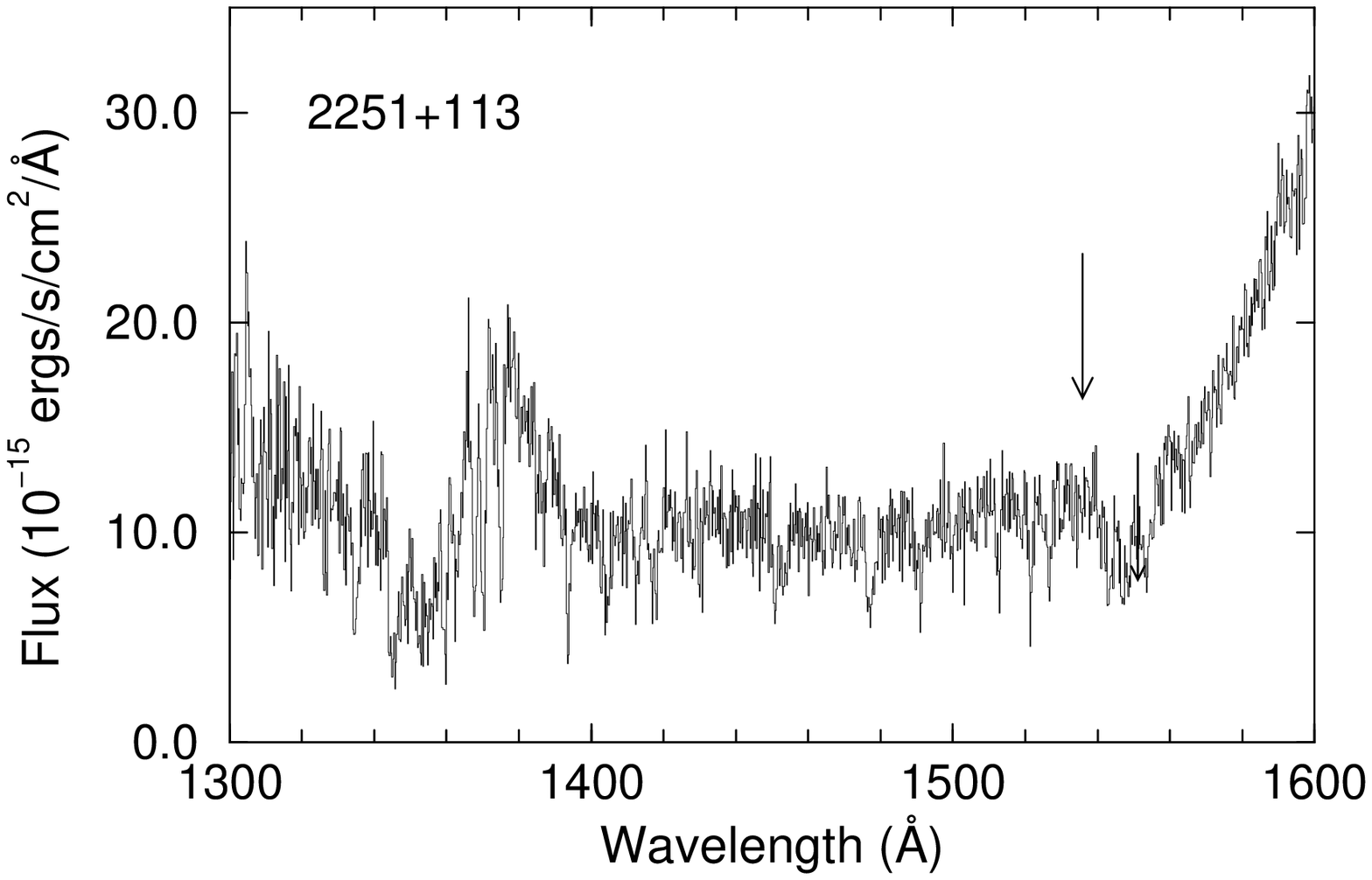}{dummyfig.eps}
\caption{{\it contd.}}
\end{figure}

\end{document}